\documentclass{emulateapj}
\usepackage{epsfig,natbib,amsmath}

\newcommand{\mic}{\,$\mu$m }
\newcommand{\micpa}{\,$\mu$m}

\newcommand{\spi}{{\it Spitzer} }
\newcommand{\spipa}{{\it Spitzer}}

\newcommand{\Msol}{M$_\odot$}

\bibpunct[]{(}{)}{;}{a}{}{,} 
 
\shorttitle{{\it Spitzer} spectro-imaging of the GRB980425 host galaxy}
\shortauthors{Le Floc'h et al..}

\begin{document}
\def\gtapp
{\mathrel{\hbox{\raise0.3ex\hbox{$>$}\kern-0.8em\lower0.8ex\hbox{$\sim$}}}}
\def\ltapp
{\mathrel{\hbox{\raise0.3ex\hbox{$<$}\kern-0.75em\lower0.8ex\hbox{$\sim$}}}}
\def\ts{\thinspace}

\title{The first Infrared study of the close environment of a long Gamma-Ray Burst\footnote{Based on observations made with {\it
      Spitzer}, operated by the Jet Propulsion Laboratory under NASA
    contract 1407.}  
}

\slugcomment{Accepted for publication in the Astrophysical Journal}

\author{Emeric~Le~Floc'h\altaffilmark{1}, Vassilis Charmandaris\altaffilmark{2,3,4},  Karl Gordon\altaffilmark{5}, 
William J. Forrest\altaffilmark{6},
Bernhard Brandl\altaffilmark{7},
Daniel Schaerer\altaffilmark{8,9},
Miroslava Dessauges-Zavadsky\altaffilmark{8} \& 
Lee  Armus\altaffilmark{10}
}
\altaffiltext{1}{Laboratoire AIM, CEA/DSM/IRFU, CNRS, Universit\'e Paris-Diderot, 91190 Gif, FRANCE}
\altaffiltext{2}{Department of Physics \& Institute of Theoretical and Computational Physics, University of Crete,  GR-71002, Heraklion, Greece}
\altaffiltext{3}{Chercheur Associ\'e, Observatoire de Paris, F-75014,
  Paris, France}
\altaffiltext{4}{IESL / Foundation for Research and Technology-Hellas, PO Box 1527, 71110, Heraklion, Greece}
\altaffiltext{5}{Space Telescope Science Institute,  3700 San Martin Dr.,  Baltimore, MD 21218, USA}
\altaffiltext{6}{Dept. of Physics \& Astronomy, University of  Rochester,  Rochester, NY 14627, USA}
\altaffiltext{7}{Leiden Observatory, Leiden University, Niels Bohrweg 2, P.O. Box 9513, 2300 RA Leiden, Netherlands}
\altaffiltext{8}{Geneva Observatory, University of Geneva, 51, Ch. des Maillettes, CH-1290 Versoix, Switzerland}
\altaffiltext{9}{CNRS, IRAP, 9 Av. colonel Roche, BP 44346, F-31028 Toulouse cedex 4, France}
\altaffiltext{10}{Spitzer Science Center, California Institute of Technology, 220-6, Pasadena, CA 91125, USA}

\begin{abstract} 
We present a   characterization of the close environment of GRB\,980425  based on 5--160\mic spectro-imaging obtained  with  \spipa. 
The Gamma-Ray Burst GRB\,980425  occurred in a nearby ($z$\,=\,0.0085) SBc-type dwarf galaxy, at a projected distance of 900\,pc from an HII~region with    
  strong signatures of Wolf-Rayet (WR) stars. While this ``WR~region" produces  less than 5\% of the $B$-band emission of the host, we find that it is responsible for
45$\pm$10\,\% of the total infrared luminosity, with a maximum contribution  reaching 75\% at 25--30\micpa. This atypical property is rarely observed among morphologically-relaxed dwarves,
  suggesting a strong causal link with the GRB event.  
The luminosity of the WR~region ($L_{8-1000\mu m}$\,=\,4.6\,$\times$\,10$^8$\,L$_{\odot}$), the peak of its spectral energy distribution at $\lesssim$\,100\mic and the 
presence of   highly-ionized emission lines 
 (e.g., [NeIII]) also reveal extremely young ($<$\,5\,Myr) star-forming activity, with a typical time-scale of only 47\,Myr to double the stellar mass already built. 
 Finally, the mid-IR over  $B$-band luminosity ratio in this region is substantially higher than  in star-forming galaxies with similar L$_{\rm IR}$, but it is 
 lower than in young dust-enshrouded stellar clusters. 
 Considering 
 the modest obscuration  measured from the  silicate features ($\tau_{9.7\mu m} \sim 0.015$), this suggests that the WR~region is dominated by 
  one or several star clusters that have either partly escaped or cleared out their parent molecular cloud.
 Combined with the properties characterizing the whole population of  GRB hosts, our results reinforce the idea that long GRBs mostly happen within or in the vicinity of   relatively unobscured  galactic regions harboring very  recent star formation.

\end{abstract}


\section{Introduction}

It is now widely believed that Long Gamma-Ray Bursts (LGRBs\footnote{Long GRBs, refered as LGRBs in this paper, are defined as Gamma-Ray Bursts with a duration greater than $\sim$\,2\,s and a  soft spectrum, as opposed to the class of the short and hard GRBs \citep{Kouveliotou93}.}) originate from the cataclismic collapse of massive stars
\citep[e.g.,][]{Woosley93,Galama98,MacFadyen99,Daigne00,Heger03,Stanek03,Meynet05} in star-forming regions of distant galaxies \citep[e.g.,][]{Fruchter99,Bloom02a,Hjorth02,Fruchter06}. They can be observed  at extremely  high redshift \citep[$z > 6$, ][]{Greiner09}  
and  because of their association with young and short-lived  stars, they are   invaluable
tools for sign-posting  the activity of star formation back to the earliest epochs of cosmic history and galaxy evolution.
 Furthermore, their optical and near-infrared afterglows can be used as transient background sources for probing  the intergalactic space and the interstellar medium of galaxies up to very large cosmological distances \citep[e.g.,][]{Vreeswijk01,Totani06}. As a result they might eventually provide unique constraints on the era of reionization and the formation of the very first structures in the Universe \citep{Lamb00}.

However, it appears that  LGRBs  may only trace regions of star-forming activity in a biased way, in the sense that their formation is more likely to occur 
in low-mass and sub-$L^*$ galaxies with low-metallicity environments \citep{Sokolov01,Fynbo03,Courty04,Courty07,Kistler09,Kocesvski09,Han10,Niino11}.
In the latest stages of their evolution, massive stars with metal-poor envelopes retain a higher angular momentum and they are less subject to mass loss. After the final collapse these physical conditions lead to the formation of  fast-rotating black holes and the subsequent accretion of material thus favors the formation of  GRBs \citep[e.g.,][]{MacFadyen99,Hirschi05}. In fact, observational evidence for  a possible relationship between LGRBs and chemically-young galaxies has already emerged from a number of studies.
On average the hosts of LGRBs are characterized by blue colors, rather low extinction and modest bolometric luminosities, as well as  diffuse and irregular morphologies \citep[e.g.,][]{Bloom02a,LeFloch03,LeFloch06,Fruchter06}. They also  exhibit  higher Ly$\alpha$ emission and larger specific star formation rates (SSFR) with respect to other  galaxies at similar redshifts \citep{Fynbo03,Christensen04}, and their oxygen abundance is lower compared to field star-forming sources with similar masses and luminosities \citep{Modjaz08,Han10}.
Furthermore, the connection between GRBs and Type~Ic supernovae has led to the idea that the progenitors of long GRBs should more likely originate from binary systems \citep{Podsiadlowski04}, which complicates further the link that exists between LGRBs and the cosmic history of structure formation.  
Better understanding  the physical conditions and the properties of the environments favoring the formation of these cosmic explosions appears, therefore, as a critical step before allowing a full exploitation of their hosts as cosmological tracers of galaxy evolution.

Follow-up of Gamma-Ray Bursts has shown that high-redshift LGRB hosts are challenging to characterize in detail, since they tend to be low-mass and sub-luminous systems
\citep[e.g.,][]{Savaglio09}. Host galaxies at very low redshift may thus provide us with some of the best cases to explore the intimate connection between the occurence of GRBs and the properties of their very close environment. For example, the now-popular GRB\,980425 was identified  at $z$\,=\,0.0085 from its apparent association with the hypernova SN\,1998bw \citep{Galama98}. With a luminosity distance of only 36\,Mpc it is currently the closest GRB  known with a confirmed redshift in the local Universe.
At this distance an angular separation of 1\arcsec \, 
on the sky corresponds to 170\,pc.  It provides therefore an excellent prototype for studying the immediate surrounding of LGRBs within their host galaxies.

The host of GRB\,980425 is identified as ESO\,184-G82 in the survey of the southern sky that was  performed by \citet{Holmberg77}.  Observations with the {\it Hubble Space Telescope\,}  (HST) showed that it is  an isolated and barred SBc-type sub-luminous galaxy (L$_{\rm B}$\,=\,0.02\,L$_{\rm B}^{*}$) harboring  a large number of active star-forming regions \citep{Fynbo00}. 
While the surface brightness of the area where GRB\,980425 was observed is not particularly large, a remarkable result is that the  GRB occurred  
 at a projected distance of only $\sim$\,900\,pc from a quite luminous  HII~region   
 characterized by very distinct properties compared to the rest of the system. This region is the brightest complex of star formation in the host and although
 it  only makes a few percent of the total luminosity
  at optical wavelengths, 
infrared (IR) imaging obtained  with the {\it Spitzer Space Telescope\,} revealed that its 5--25\mic spectral energy distribution (SED) is characterized by a  steeply-rising contribution with respect to 
 the whole SED of the galaxy  \citep{LeFloch06}. At 24\mic  
 it is responsible for a fraction as large as $\sim$\,75\% of the monochromatic luminosity of the entire system.   Among low-luminosity star-forming disks and morphologically-evolved dwarf galaxies, such a high contribution of extra-nuclear star formation  is rarely observed.
It makes  the host of GRB\,980425 an atypical object of the local Universe.

While the stellar mass already assembled in this  region is
 low ($\sim$\,6\,$\times$\,10$^{6}$\,\Msol, \citealt{Michalowski09}), 
long-slit spectroscopy   revealed the presence of numerous  Wolf-Rayet (WR) stars  as well as prominent H$\alpha$ and H$\beta$ emission lines with very large equivalent widths \citep{Hammer06}.   A metallicity of Z\,$\sim$\,0.5\,Z$_{\odot}$ was inferred from the effective temperature of the ionized gas \citep{Hammer06} and the estimated  star formation rates (SFR)  vary between 0.06 and 0.3\,\Msol\,yr$^{-1}$ \citep{LeFloch06,Hammer06,Christensen08,Michalowski09}. 
This implies a large  SSFR  and suggests
 that this environment (denoted as the ``WR~region'' hereafter)  is  experiencing a young and vigorous starburst. Finally, integral field spectroscopy performed by \citet{Christensen08} with the VIMOS instrument revealed that the ionizing conditions in this area are markedly different from those observed in the other HII~regions found in the host. With a particularly strong [OIII]/H$\beta$ line ratio ($\gtapp$\,5), the ionized gas properties observed in the WR~region are typical of very young starbursts at sub-solar metallicities,  and they also
appear very similar to the properties characterizing the  more distant GRB host galaxies as  a whole.
Along with the detection of  WR~stars  considered as very strong candidates of collapsar progenitors, the emission line diagnostics suggest therefore an intimate connection between the characteristics of this source and the physical mechanisms triggering the formation of LGRBs in the Universe.   \citet{Hammer06} even proposed that the WR~region may have been the true birth place of the  GRB\,980425 progenitor.  The latter would have been dynamically ejected  from this environment, to finally produce a collapsar after traveling  at velocities $\sim$\,200--300\,km\,s$^{-1}$ during a few Myrs through the interstellar medium (ISM) of the galaxy.

Following our previous characterization of the GRB\,980425 host based on 4.5\micpa, 8\mic and 24\mic broad-band imaging  \citep{LeFloch06} we performed additional IR follow-up with the {\it Spitzer Space Telescope\,} in order to further constrain  the spectroscopic mid-IR properties as well as the far-IR luminosity of the WR~region and the host galaxy. Here we report on the 5--35\mic spectrum that we obtained with the {\it Infrared Spectrograph for Spitzer\,} \citep[IRS, ][]{Houck04a} as well as the 70\mic and 160\mic imaging obtained with the {\it Multi-band Imager and Photometer for Spitzer\,}  \citep[MIPS, ][]{Rieke04}. Along with the host of GRB\,031203 ($z$\,=\,0.1055) that was also observed with the IRS by \citet{Watson11}, ESO\,184-G82 represents one of the very few identifications of LGRB host galaxies
located close enough  to allow detailed characterization in the thermal infrared. Our observations are described in Sect.\,2, while in Sect.\,3 we present the broad-band photometry and the spectroscopic measurements  derived from the \spi data. 
In Sect.\,4 we derive the total IR SEDs and star formation rates of the host and the WR~region, and in Sect.\,5 we constrain in more detail their mid-IR spectroscopic properties. We discuss our results in Sect.\,6 and we  present our conclusions in Sect.\,7.  Throughout this work, physical quantities are computed assuming H$_{0}$\,=\,70 km\,s$^{-1}$ as well as the Initial Mass Function (IMF) from \citet{Salpeter55}.

\vskip 1cm

\section{Observations and Data Reduction}

\subsection{Far-Infrared imaging}

The host galaxy of GRB\,980425 was imaged at 70\mic and 160\mic with the MIPS instrument as part of a  \spi  Cycle~3 General Observer program (PID: 30\,251, AOR keys 24966656 \& 24966400). The observations were carried out in 2008 using the MIPS ``Small Scale Photometry'' mode. With this operating mode a set of 10~frames is taken for each observing cycle and a dither is automatically performed around the targeted source. We respectively obtained 5 and 12~cycles at 70\mic and 160\micpa, which resulted in a total on-source integration time of 545\,s at 70\mic and 250\,s at 160\micpa.  

For a very modest amount of additional requested time we also performed a new observation of the GRB\,980425 host at 24\micpa, with the goal of checking if some fraction of the luminosity of the bright 24\mic point source  detected in our previous data \citep{LeFloch06} could be associated to the emission from a transient object (e.g., supernova). A single cycle of 10~frames was obtained, leading to a total integration time of 165\,s on source.

Each individual image was processed with the most recent version (v.3.1) of the MIPS Data Analysis Tool \citep[DAT,][]{Gordon05}. This update  includes additional processing steps with respect to the first version of the DAT initially released. In particular a pixel-dependent background subtraction was applied to each individual frame before co-adding the images. We used calibration factors of 702\,(MJy\,sr$^{-1}$)/(ADU) and 41.7\,(MJy\,sr$^{-1}$)/(ADU)  at 70\mic and 160\mic respectively \citep{Gordon07,Stansberry07}. The pixel size of the final mosaics was set to half  the pixel size of the MIPS detectors and the absolute astrometric registration was tied to the \spi astrometry of the raw data as provided by the \spi  Science Center.  The  images obtained at the 3 MIPS wavelengths are presented in Fig.\,1. For reference  they are compared with the image of the GRB\,980425 host galaxy obtained at 4.5\mic from our first \spi  observations \citep{LeFloch06}.

\vskip .4cm

\begin{figure*}[htpb]
  \epsscale{1.1}
  \plotone{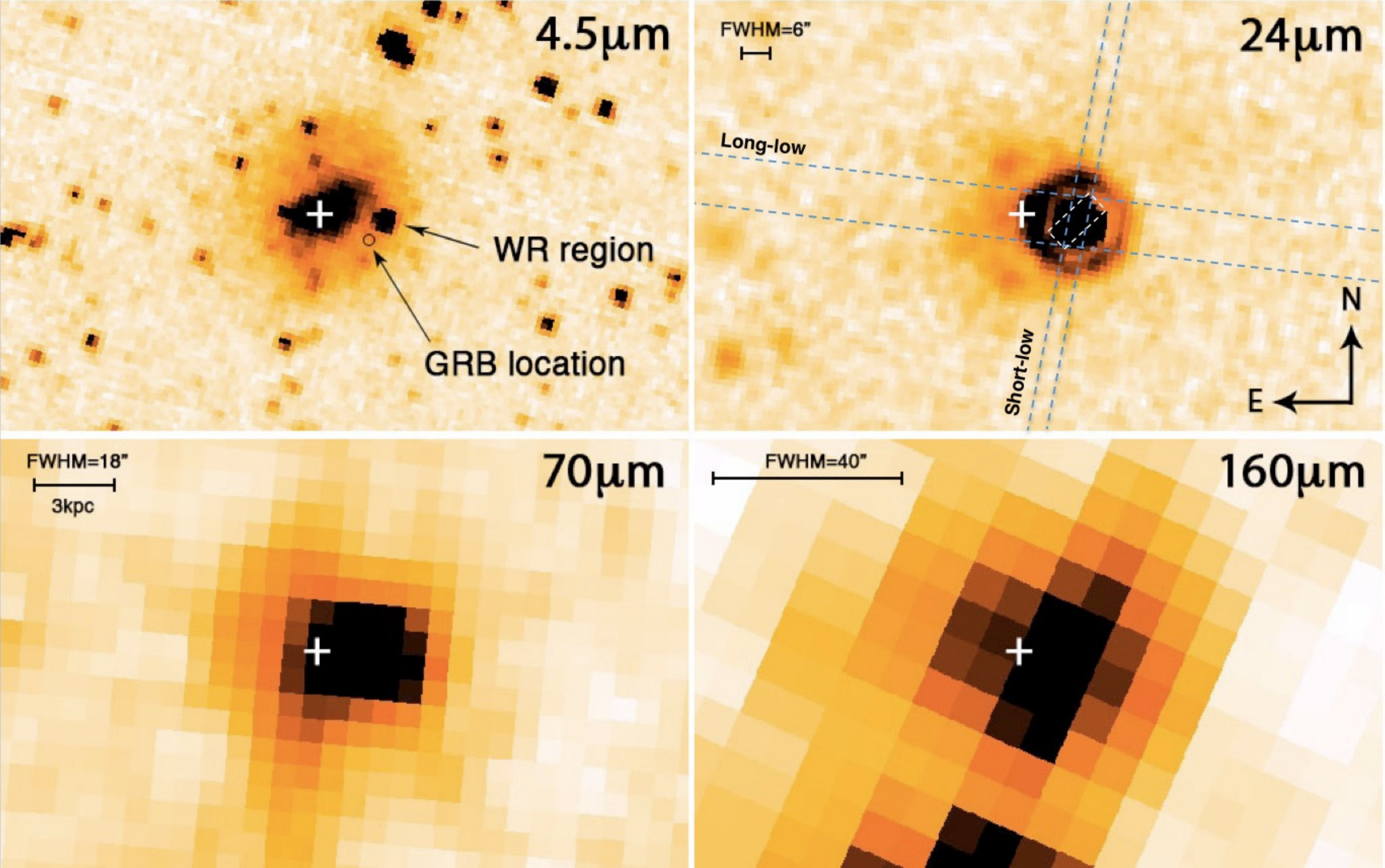}
  \caption{The host galaxy of GRB\,980425 imaged by \spi  at 4.5\mic (top-left), 24\mic (top-right), 70\mic (bottom-left) and
160\mic (bottom-right). The four panels have the same field of view, North is to the top and East is to the left. The position of the previously identified Wolf-Rayet (WR) region  and the location where the GRB was observed  (see Sect.\,1)  are shown in the 4.5\mic image. The physical scale is given in the bottom-left panel while the beam size of the instrument is indicated for each of the MIPS observations.   The 24\mic image also illustrates the position of the slit apertures for the IRS observations performed with the low  (blue dashed lines) and high resolution (white dashed line) modules.  In each panel the white 'cross' symbol (+) pinpoints the center of mass of the galaxy defined from the 4.5\mic image, which shows that the strong emission detected at 70\mic and 160\mic is shifted toward the WR~region.}
\label{fig:4panels}
\vskip .2cm
\end{figure*}

\subsection{Mid-Infrared spectroscopy}
\label{sec:irs}

The mid-IR spectroscopic follow-up of the WR~region was
performed as part of the  IRS Guaranteed Time Observation (GTO)
program on 19 April 2006 (AOR key: 12246528). The target was first acquired using the red
peak-up (RPU) camera. During this acquisition  an image of the source at 22\mic was obtained 
 in Double Correlated Sampling (DCS) mode to locate the
mid-IR centroid of the target, and then offset to the appropriate slit
(see the Spitzer Observers Manual for more details).  Using the Short-Low
(SL, 5.2-14.5\micpa) and Long-Low (LL, 14.0-38.0\micpa) modules of
the  IRS, we obtained a 5$\sim$38\mic low resolution
spectrum of the WR~region. The total on-source exposure time was 244\,s (2~cycles of
60\,s) per SL order and 126\,s (2 cycles of 30\,s) per LL order. In
addition, higher resolution observations were obtained with the IRS Short-High (SH, 9.9-19.6\micpa) 
for a total on-source time of 975\,s (4~cycles of 120\,s).   For each  mode of our IRS observations the orientation of the slit aperture with respect to the GRB\,980425 host galaxy  is illustrated in Fig.\,1 (top-right panel).

We begun analyzing the observations by processing the data with
the {\em Spitzer} Science Center pipeline
(version~15.5) and the extraction of the signal was carried out with
 the Spectral Modeling, Analysis, \and
Reduction Tool \citep[SMART Ver. 5.5.1,][]{Higdon04b}. In the case of the low-resolution data though, the final reduced spectrum later used in this paper was retrieved from
the recently published {\it Cornell AtlaS of Spitzer IRS Sources\,} \citep[CASSIS, ][]{Lebouteiller11b}, which provided us with an  improved and updated processing of the data.

In brief, the IRS data reduction starts from
the intermediate pipeline products (droop files), which only lack
stray light and flat field correction. Individual pointings to each
nod position of the slit were co-added using median averaging. For SL
and LL spectra, we considered the difference of the two nod positions to
remove the contribution of the background. Then we extracted the
spectra for each nod position using a variable width aperture, which
scales the extraction aperture with wavelength to recover the same
fraction of the diffraction limited instrumental Point Spread Function
(PSF). The data from SH were extracted using the full slit extraction
method from the median of the combined images.  Since no sky
(off--position) measurements were taken for the high resolution module, the contribution of the sky
emission was not subtracted from SH spectra.  Then the spectra were
flux calibrated by multiplication with the Relative Spectral Response
Function (RSRF), which was created from the  IRS standard stars
$\alpha$ Lac for SL and LL and $\xi$ Dra for SH, for which accurate
templates were available (Cohen et al. 2003). We built our RSRFs by
extracting the spectra of the calibration stars in the exact same way
as the science target, and dividing the stellar templates by the
extracted stellar spectra. We produced one RSRF for each nod position
in order to avoid systematic flat field errors.  The signal difference
between the nod positions provide the error estimates.  Finally, the
flux calibrated spectra of each order of the low-resolution modules
were scaled to the first order of LL (LL1, 20--36$\mu$m), which was used to
define the overall continuum of the source. 

\begin{figure}[htpb]
  \epsscale{1.2}
  \plotone{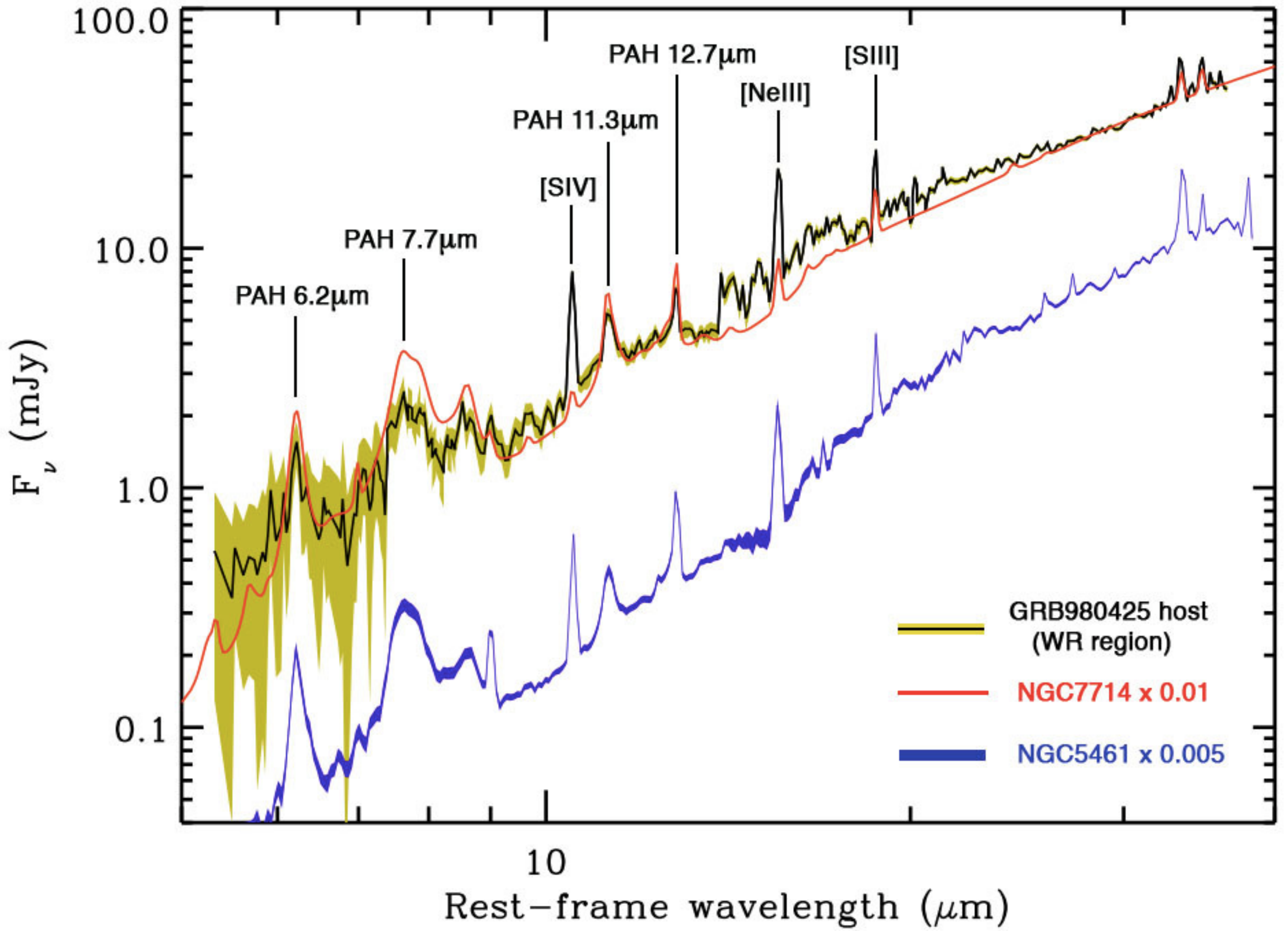}
\vskip .3cm
  \caption{The IRS low-resolution spectrum of the WR~region in the host galaxy of GRB\,980425 (black solid line), compared with the spectral energy distribution of the starburst galaxy NGC\,7714 \citep[, red solid line]{Brandl04} and the SED of the HII~region NGC\,5461 in M\,101 \citep[, blue solid curve with thickness indicating the  1$\sigma$ uncertainty]{Gordon08}.  The light shaded area around the GRB host spectrum represents the 1$\sigma$ error bar estimated for each resolution element. Similar to NGC\,7714 and NGC\,5461 the WR~region exhibits a steeply rising continuum and bright PAH features, but also quite prominent ionic emission lines.} 
\label{fig:plot_spec}
\vskip .2cm
\end{figure}

The final low-resolution IRS spectrum of the WR~region is shown in Figure~\ref{fig:plot_spec}. Given the star-forming nature of  GRB host galaxies  we  compared it with the IRS observations of the starburst galaxy NGC\,7714 \citep{Brandl04} and the super-luminous HII~region NGC\,5461 in the spiral galaxy M\,101 \citep{Gordon08}. The three SEDs exhibit striking similarities but also noticeable differences. First, they  are all characterized by a steeply-rising continuum of hot dust emission, and the Polycyclic Aromatic Hydrocarbons (PAH) features commonly seen in star-forming galaxies \citep{Laurent00,Brandl06,Smith07} are also clearly apparent. These PAHs as well as the continuum emission originate from dust grains stochastically heated by the radiation field and by the UV photons  from young populations of massive stars. Their luminosity correlates  very well with the total IR luminosity measured between 8\mic and 1000\mic \citep{Brandl06}  and they are therefore considered as a fairly accurate and extinction-free tracer of the activity of star formation in galaxies \citep[e.g.,][]{Roussel01,Calzetti07,Diaz_Santos08}.

However we note that the 7.7\mic PAH feature in the GRB host is substantially weaker than in NGC\,7714 and NGC\,5461, while the main other PAHs detected at 6.2\micpa, 11.3\mic and 12.7\mic have roughly the same relative strengths in the three spectra. Furthermore, the mid-IR spectral slopes are almost identical between the WR~region and NGC\,7714, while the mid-IR spectrum of the HII~region NGC\,5461 seems to rise even more rapidly at the longest wavelengths.
Finally the IRS spectrum of the WR~region reveals  the presence of prominent fine-structure emission lines such as  [SIV]-10.51\micpa,
 [NeIII]-15.55\mic  and
[SIII]-18.71\micpa, which are also detected in NGC\,5461 but either absent or much fainter in NGC\,7714. In the following sections we will quantitatively analyze the relative strengths of the different features characterizing the WR~region of the GRB\,980425 host. In particular  we will further explore how it differs from what has been so far observed in the mid-IR spectra of star-forming galaxies.

\section{Measured quantities}

\vskip .2cm

\subsection{MIPS broad-band photometry}
\label{sec:photo}

This section describes the broad-band photometry measurements obtained from our MIPS imaging. All the measured fluxes are summarized in Table\,\ref{table:prop_IR}.

\begin{deluxetable*}{lccccc}
\tablenum{1}
\footnotesize
\tablecaption{Infrared  photometry  and global properties of the WR~region and the GRB host}
\tablewidth{0pc}
\tablehead{
\colhead{}  & & & \colhead{WR~region} &  \colhead{GRB\,980425 host (total)} &  \colhead{Reference}
}
\startdata
$F_{8\mu m}$                    &       &              &  1.8$\pm$0.1\,mJy    & 11.9$\pm$0.3\,mJy  &  \citet{LeFloch06} \\ 
$F_{24\mu m}$                    &        &             & 19.5$\pm$2.0\,mJy   & 26.2$\pm$1.3\,mJy &  Sect.\,3.1.3   \\ 
$F_{70\mu m}$                     &         &             & 140$\pm$25\,mJy   & 230$\pm$30\,mJy  &  Sect.\,3.1.1 \\ 
$F_{160\mu m} $                   &        &             & $-$    & 615$\pm$200\,mJy  &  Sect.\,3.1.2\\ 
Log($L_{IR}/L_\odot$)                     &  &                & 8.66$\pm$0.04 & 9.01$\pm$0.07 & Sect.\,4.1 \\ 
$SFR$                     &  &                & 0.12\,\Msol\,yr$^{-1\,(a) }$ & 0.16\,\Msol\,yr$^{-1\,(b)}$ & Sect.\,4.2 \\ 
$\tau_{9.7\mu m}$ & & & 0.015 & $-$ & Sect.\,5 \\
\enddata
\tablenotetext{a}{Using the calibration from \citet{Bell05}.}
\tablenotetext{b}{Using the calibration from \citet{Calzetti07}.
\\
}
\label{table:prop_IR}
\end{deluxetable*}

\vskip .2cm
\subsubsection{MIPS-70\micpa}

As one can see from Figure\,\ref{fig:4panels} the host of GRB\,980425 is clearly detected at 70\micpa, and in spite of the modest spatial resolution of the data, low surface brightness  emission can also be observed up to a distance of $\sim$\,35\arcsec \, from the central region of the galaxy. To determine the total 70\mic  flux density of the host we  first estimated the   sky  background in the image by taking the median value of  the pixels in  the outskirt regions   up to $\sim$\,100\arcsec \, from the center of the source.  We then subtracted this sky level to each pixel of the mosaic and we measured the photometry  inside a circular aperture with a diameter of 70\arcsec \, centered on the nucleus of the host. The photometric uncertainty was estimated from the pixel-to-pixel variations measured over the sky. We derived a total flux of $S_{\rm 70\mu m}$\,=\,230$\pm$30\,mJy.

\begin{figure}[htpb]
  \epsscale{1.1}
 \plotone{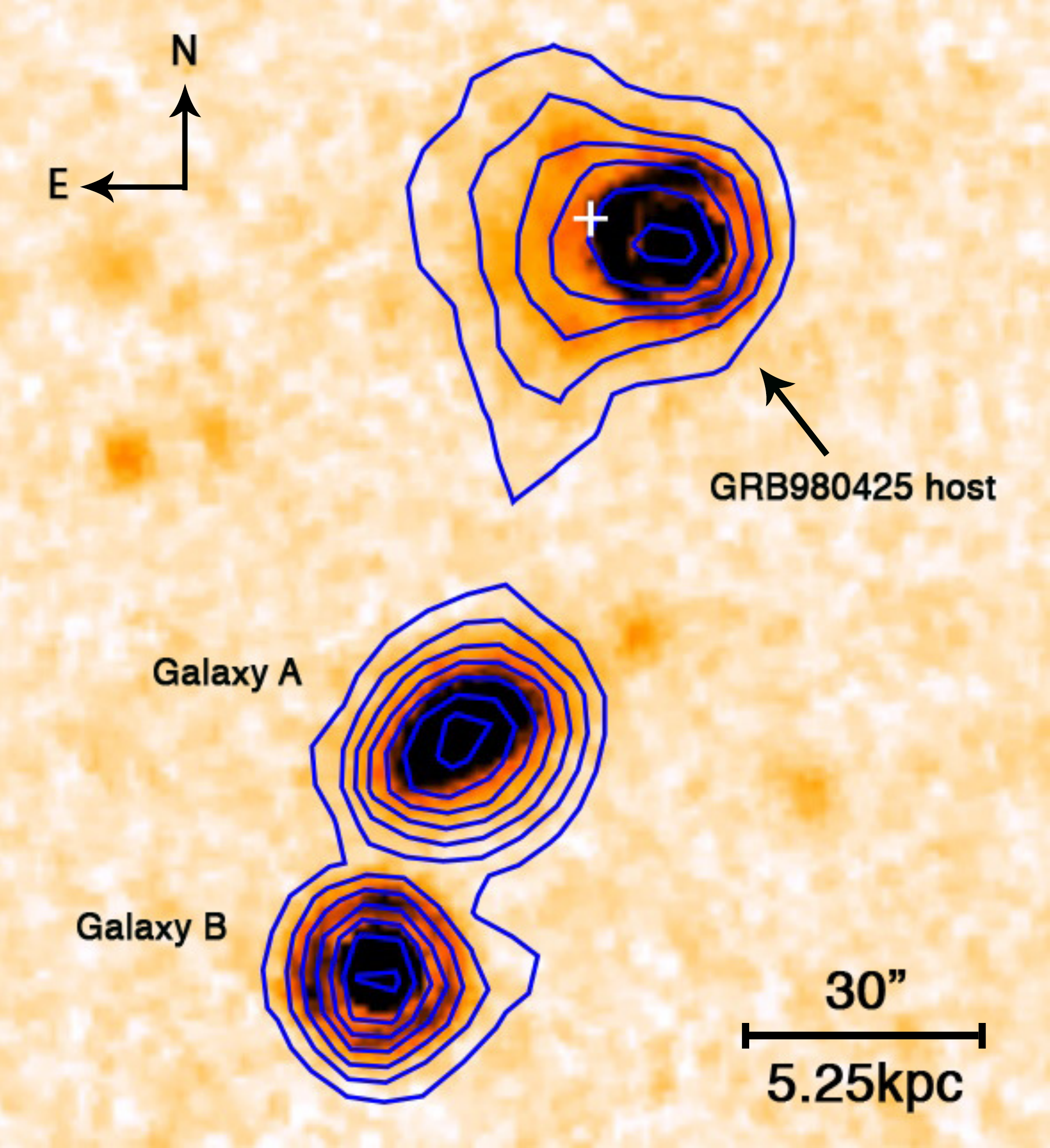}
  \caption{The surrounding of the GRB\,980425 host galaxy imaged at 24\micpa, superimposed with the contours of the 70\mic emission. The GRB host is visible to the North while Galaxy~A and Galaxy~B identified by \citet{Foley06} are also clearly detected at both wavelengths to the South East. The contours scale linearly from 1.7 to 15\,MJy\,sr$^{-1}$. As in Figure~\ref{fig:4panels} the white 'cross' symbol (+) indicates the center of mass of the GRB host galaxy. In spite of the large FWHM of the far-IR MIPS PSF the peak of the 70\mic emission detected toward the host of GRB\,980425  clearly coincides  with the bright 24\mic point source that corresponds to the WR~region.}
\label{fig:contours_70mic}
\end{figure}

Furthermore, the peak of the 70\mic emission is clearly shifted from the center of mass of the host defined from the 4.5\mic stellar emission, and  there is strong evidence that the bulk of the 70\mic luminosity  originates from the WR~region that also dominates the mid-IR luminosity of the  galaxy.  Figure\,\ref{fig:contours_70mic} shows a larger field of view of the MIPS-24\mic image of the GRB\,980425 host, overlaid with the contours of the emission detected at 70\micpa. 
We see that the peak of the 70\mic emission
coincides with the very bright 24\mic point source lying
 in the South-West spiral arm at 12\arcsec \, from the center of the host galaxy. 

  Although this distance is slightly smaller than the FWHM of the MIPS-70\mic PSF (18\arcsec), 
we believe that the significance of this shift is  robust given the high signal to noise of the 70\mic emission detected toward the GRB host. As described in the next paragraph the peak of the 70\mic emission originates from a point source detected with a signal to noise $S/N \sim 5.6$. The positional 1$\sigma$ uncertainty associated with this point source can be estimated as $0.6 \times FWHM \times (S/N)^{-1} = 6.75\arcsec$ (see Appendix~B of \citealt{Ivison07}), which is  twice smaller than the measured offset.
Also, such a shift with respect to the nucleus of the galaxy could not be artificially produced by a misalignment between the 24\mic and 70\mic mosaics given the astrometric calibration of the \spi data (1$\sigma \sim$\,1\arcsec \,  rms). In fact there are two other galaxies detected to the South of the GRB host  (see Figure\,\ref{fig:contours_70mic}) and their 70\mic emission appears to be very well centered on the peak of their 24\mic counterpart.

We compared the two-dimensional profile of the 70\mic detection of the GRB\,980425 host with the MIPS-70\mic PSF. In spite of the lower spatial resolution of the MIPS data at this wavelength  we found that there is  a spatially-resolved low surface bightness component   extending to the East of the peak of the 70\mic emission.  This component is observed over the whole extent  
 of the galaxy and  it is clearly apparent from the asymmetrical shape of the contours displayed in Figure\,\ref{fig:contours_70mic} (note however that the small extension to the South comes from a data reduction artifact produced along one of the two dimensions of the 70\mic pixel array). 
We thus decomposed the 70\mic image of the GRB\,980425 into 
 a point source centered on the peak of the 70\mic emission 
and the contribution from the spatially-extended regions
using the PSF fitting routine of the DAOPHOT package \citep{Stetson87}. We found that the point source contributes to 140$\pm$25\,mJy,  which shows that  
a fraction as large as $\sim$\,60$\pm$10\% of the 70\mic emission of the GRB\,980425 host  originates  from the WR~region.

\vskip .3cm
\subsubsection{MIPS-160\micpa}
\label{sec:mips160}

At 160\mic the GRB\,980425 host  is also clearly detected in our data  (Figure~\ref{fig:4panels}). However the  FWHM of the MIPS PSF at these wavelengths is comparable to the angular extent of the galaxy on the sky  ($\sim$40\arcsec), and given the modest signal to noise of the observations it was not possible to confirm the presence of a spatially-extended component  similar to what we found in the 70\mic image.  The total flux density was therefore estimated with aperture photometry  in a radius of 40\arcsec, and a correction taking account of the PSF emission outside of this aperture was applied following the prescriptions described by \citet{Stansberry07}. We obtained a total flux of  $S_{\rm 160\mu m}$\,=\,615$\pm$200\,mJy. 
Given the relatively small size of the 160\mic pixel array as well as the contamination from the galaxy detected to the South of the GRB host (referred as Galaxy~A by \citealt{Foley06}, see also Figure~\ref{fig:contours_70mic}), only a very small number of pixels were available to estimate the sky background in the image. This explains the  large photometric uncertainty affecting our flux estimate.

Given the low angular resolution
 at these long wavelengths
it was not possible to perform a spatial decomposition similar to the one described above for the 70\mic data. In particular, we could not determine the fraction of energy powered  by the point source that dominates the monochromatic luminosity at shorter IR wavelengths. However, Figure\,\ref{fig:4panels} reveals that the peak of the 160\mic emission is again  shifted to the South-West with respect to the center of the galaxy. 
Within the astrometric uncertainties  it coincides with the peak of emission detected at 24\mic and 70\micpa, which suggests
that the luminous HII region detected in the mid-IR is also responsible for a very large fraction of the total far-IR emission of the galaxy\footnote{The host of GRB\,980425 is the only object covered by the field of view of the 160\mic observations. Therefore it was not possible to independently confirm the astrometry with other field galaxies as we did  at 70\mic and we had to rely on the \spi astrometric registration which is accurate to  1\arcsec.}.

\vskip .6cm
\subsubsection{MIPS-24\micpa}
\label{sec:mips24}

The new 24\mic image of the GRB\,980425 host galaxy  obtained in 2008 as part of this program looks very similar to the  observations  performed by MIPS/\spi in 2004 \citep{LeFloch06}. Most of the emission originates from a single point-like source that appears to coincide with the WR~region, while a couple of other HII regions as well as a more diffuse component are also detected toward  the nucleus and in the spiral arms of the host.
We measured the total flux of the galaxy and the luminosity of the WR~region using a PSF fitting decomposition and  the same approach followed for the 70\mic data. We found that the unresolved component contributes to a level of 19.5$\pm$2.0\,mJy, while the total emission of the galaxy reaches $S_{\rm 24\mu m}$\,=\,26.2$\pm$1.3\,mJy. Within the error bars these results are fully consistent with the 24\mic photometry of \citet{LeFloch06}. The lack of temporal variations shows that if some dust heating transient emission produced after the GRB explosion has contributed to the mid-IR emission of the galaxy, this emission must either be very faint (i.e., $\ltapp$10\% of the total luminosity) or decrease very slowly with time.

\vskip .6cm

\subsection{Mid-Infrared spectroscopic features}
\label{sec:pahfit}

\begin{figure}[htpb]
  \epsscale{1.15}
\plotone{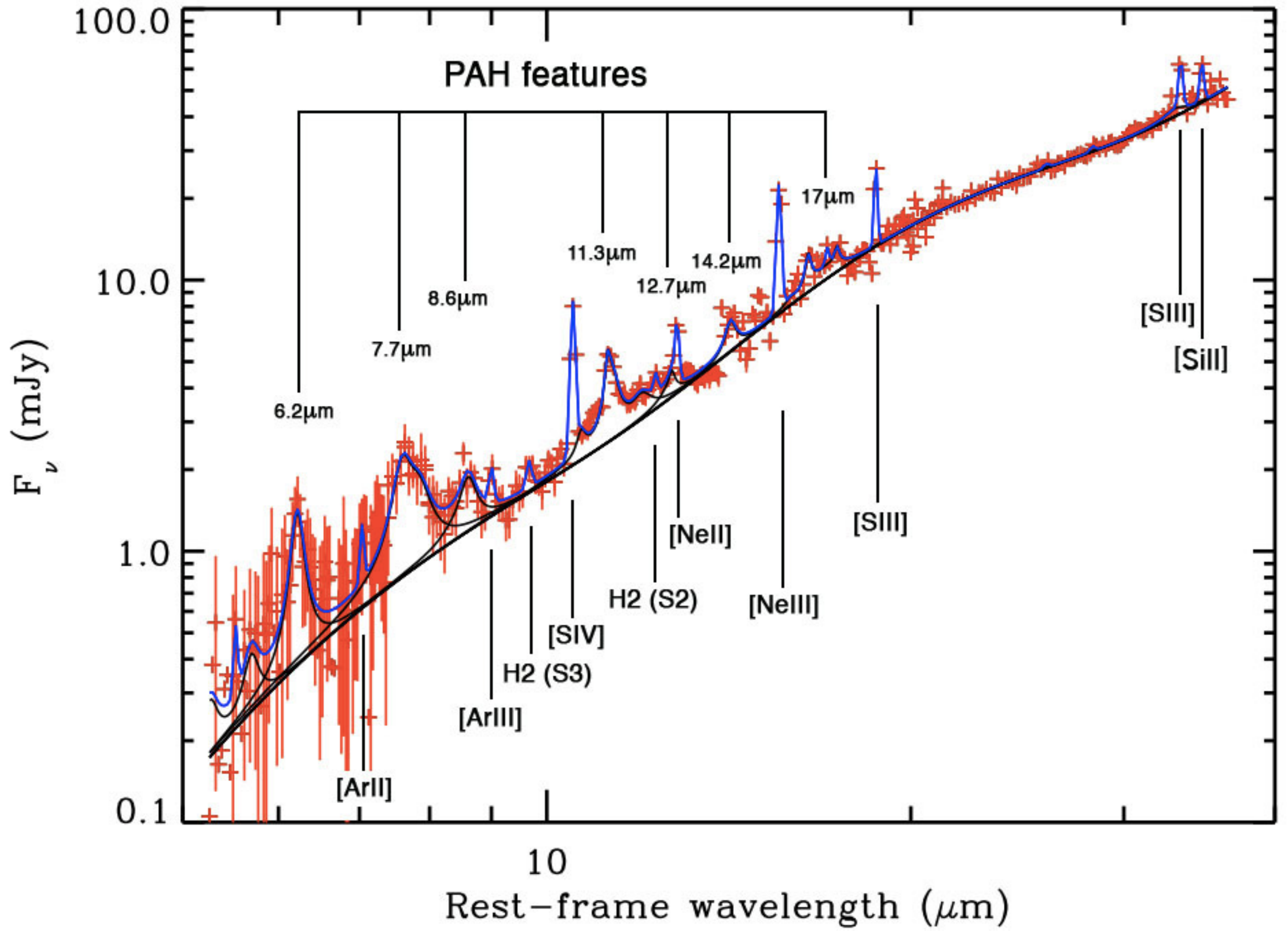}
  \caption{Decomposition of the IRS spectrum observed in the WR~region of the GRB host, obtained with the PAHFIT tool of \citet{Smith07}. The measured SED  is illustrated with the red data points and the associated error bars, while the total SED reconstructed from the model is shown with the blue solid line. The underlying continuum and the modeling of the main PAH features at 5.3, 6.2, 7.7, 8.6, 11.3, 12.7, 14.2 and 17\mic are represented by the black solid curves. The position of the different ionic lines identified by the fit are also indicated.}
\label{fig:pahfit}
\vskip .2cm
\end{figure}

Numerous emission lines are clearly visible in the IRS~spectroscopic observations of the WR~region (Figure\,\ref{fig:plot_spec}). They are commonly observed in the spectral energy distribution of star-forming galaxies and they provide valuable information on their activity of star formation as well as on the strength and the hardness of their radiation field. To constrain the luminosity of these features we analyzed our  low-resolution spectrum with the PAHFIT tool developed by \citet{Smith07}. This code was specifically designed for decomposing  \spi IRS spectra  into the contribution of the different components that characterize galaxy mid-IR SEDs. These are
 the hot dust continuum produced by the Very Small Grains (VSG), the broad-band PAH features, the narrow ionic emission lines and the effect of silicate absorption at 9.7\mic and 18\micpa. The result of our decomposition is displayed in Figure~\ref{fig:pahfit}. Given the  large number of individual features  identified with PAHFIT, only the spectral profile corresponding to the main PAHs is represented, along with the shape of the underlying continuum (thick solid line). Similarly, only the position of the most prominent forbidden lines is labeled on the plot, and the total SED reconstructed from the model is shown as a blue solid line.
 
\begin{figure}[htpb]
  \epsscale{1.1}
\plotone{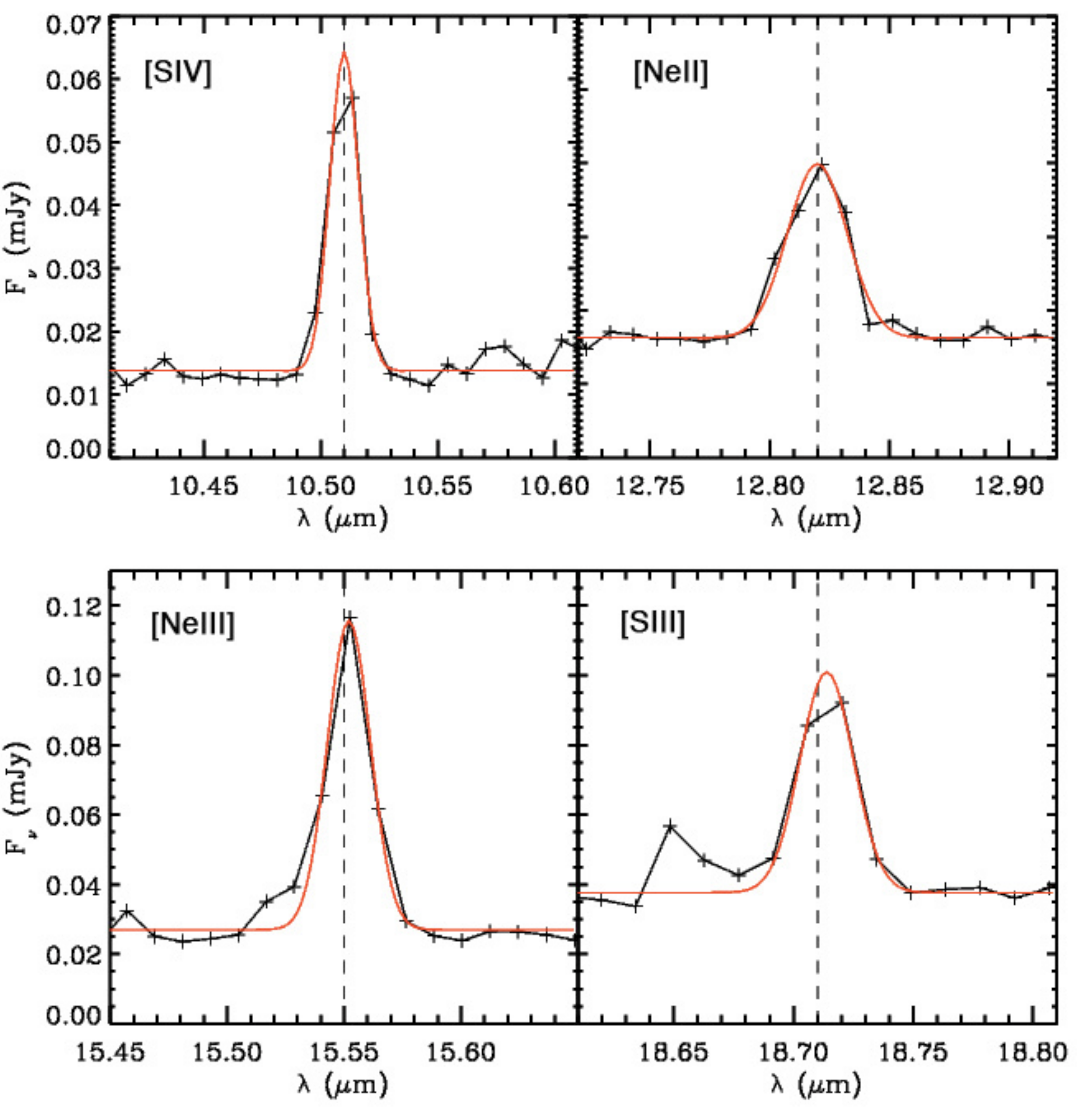}
  \caption{
  A close-up view of the profiles of the [SIV], [NeII], [NeIII] and [SIII]  ionic lines observed with the IRS Short-High  module. In each panel the observed data corrected for the redshift of the source are shown with the black ticked-solid curve while the model of the emission profile is performed with a  gaussian fit illustrated in red; the expected position of the forbidden line is indicated by the vertical dashed line.}
\label{fig:hires}
\vskip .2cm
\end{figure}

 The narrow ionic lines observed between 10\mic and 19\mic 
  are also clearly visible in the high-resolution spectrum that we obtained with IRS. A more accurate estimate of their luminosity was therefore obtained by fitting a gaussian function to their emission profile derived from the SH data. This is illustrated in  Figure\,\ref{fig:hires}, which shows the SH IRS observations obtained for the [SIV], [NeII], [NeIII] and [SIII] ionic lines. Since no sky subtraction could be performed for these data though, the underlying continuum and the equivalent widths (EW)  of these features were estimated from the continuum of the low-resolution spectrum as modeled with PAHFIT.   Our measurements are given in Table~\ref{table:features} along with the fluxes and the equivalent widths of the main PAHs and ionic lines measured in the low-resolution data. For the strongest and isolated PAHs (i.e., 6.2\micpa, 7.7\micpa, 8.6\mic and 11.3\micpa), we indicate the results obtained with the global PAHFIT decomposition but we also provide the measures that we derived with a local fit of the continuum underlying each individual feature using a spline function. The latter approach has  been commonly used in the literature to characterize the mid-IR spectra of star-forming galaxies. 
It usually leads to lower values than obtained with PAHFIT, since PAHFIT accounts for the full extent of the PAH wings. 

\begin{deluxetable}{lccccc}
\tablenum{2}
\footnotesize
\tablecaption{Mid-IR Emission Features}
\tablewidth{0pc}
\tablehead{
\colhead{Feature}  & & $\lambda$ & & \colhead{Flux} &  \colhead{EW}\\ 
 & & \colhead{($\mu$m)} & &  \colhead{$\times10^{-21}$ (W\,cm$^{-2})$} & \colhead{($\mu$m)}
}
\startdata
PAH$^a$                    &  & 6.2     &              & 3.14$\pm$0.33   (1.7$\pm$0.07)   & 0.98 (0.38)   \\ 
PAH$^a$                    &  & 7.7     &             & 9.89$\pm$1.39   (3.7$\pm$0.5)   & 2.12 (0.43)  \\ 
PAH$^a$                    &  & 8.6      &             & 1.98$\pm$0.23   (0.56$\pm$0.2)   & 0.34 (0.06)  \\ 
PAH$^a$                    &  & 11.3     &             & 4.03$\pm$0.20 (1.7$\pm$0.3)     & 0.49 (0.15) \\ 
PAH$^b$                    &  & 12.7      &            & 1.16$\pm$0.23      & 0.12  \\ 
PAH$^b$                    &  & 17          &        & 7.58$\pm$0.75      & 0.58   \\
${\rm[ArII]}$$^b$        &  & 7.01   &	& 0.18$\pm$0.16            & 0.05\\
${\rm[ArIII]}$$^b$       &  & 9.01 &	& 0.19$\pm$0.08            & 0.04\\
${\rm[SIV]}$$^c$        &  & 10.51&	& 2.01$\pm$0.18            & 0.40\\
${\rm H2(S2)}$$^c$   &  & 12.33&	& 1.35$\pm$0.07            & 0.22\\
${\rm[NeII]}$$^c$       &  & 12.82&	& 1.48$\pm$0.11            & 0.23\\
${\rm[NeIII]}$$^c$      &  & 15.55&	& 2.68$\pm$0.14            & 0.34 \\
${\rm[SIII]}$$^c$        &  & 18.71&	& 1.57$\pm$0.12            & 0.16 \\
${\rm[SIII]}$$^b$        &  & 33.53  &      & 1.90$\pm$0.09           & 0.17 \\
${\rm[SiII]}$$^b$        &  & 34.86&	& 1.44$\pm$0.09           & 0.13
\enddata
\tablenotetext{a}{Main values result from PAHFIT measurements in the low resolution data. Values in parenthesis give another estimate independently obtained using a local spline-type fitting of the continuum.}
\tablenotetext{b}{Measured with the IRS low resolution module using PAHFIT.}
\tablenotetext{c}{Measured with the IRS high resolution module, except for the continuum which was derived from the low resolution data.
\\
}
\label{table:features}
\end{deluxetable}

\section{Total IR SEDs and star formation rates}

\subsection{Infrared SED fitting}
\label{sec:ir_sed}

We constrained the spectral energy distributions of the GRB host galaxy and the WR~region over the full infrared wavelength range by fitting the \spi  broad-band photometry with the empirical libraries of galaxy  templates published by  \citet{Chary01}, \citet{Dale02} and \citet{Lagache04}, as well as with the physical SEDs derived from radiative transfer modeling by \citet{Siebenmorgen07}.  Between these different libraries the SEDs mostly vary in the relative strength of the PAH features with respect to the hot dust continuum, as well as in the temperature and the emissivity of the cold dust component shaping the peak of the SED in the far-IR. In the library of  \citet{Siebenmorgen07} the  SEDs also depend on the size of the star-forming region responsible for the IR emission, and they are given for radii of 0.35, 1, 3, 9 and 15\,kpc. We only considered sizes of 0.35\,kpc and 3\,kpc for the WR~region and the whole galaxy, respectively.

To obtain the best possible constraints  we combined the MIPS fluxes presented in Sect.\,\ref{sec:photo} with the IRAC-8\mic photometry already published by \citet{LeFloch06}. The fitting was performed separately for each library, 
using the code {\it LePhare\,} \citep{Arnouts99,Ilbert06}. Although most of the IR SED templates from the aforementioned libraries vary as a function of the total IR luminosity (but see \citealt{Dale02} for a dependence on dust temperature), their normalization was kept as a free parameter and the best templates were derived from a basic $\chi^2$~minimization of the fit.
In the case of the WR~region, we did not include the photometry at 160\mic since at this wavelength  we were unable to separate its contribution from the emission of the host (see Sect.\,\ref{sec:mips160}). We   checked however that the best fits obtained for the WR~region did not exceed the total flux measured for the GRB host galaxy at 160\micpa.

\begin{figure*}[htpb]
  \epsscale{1.1}
\plotone{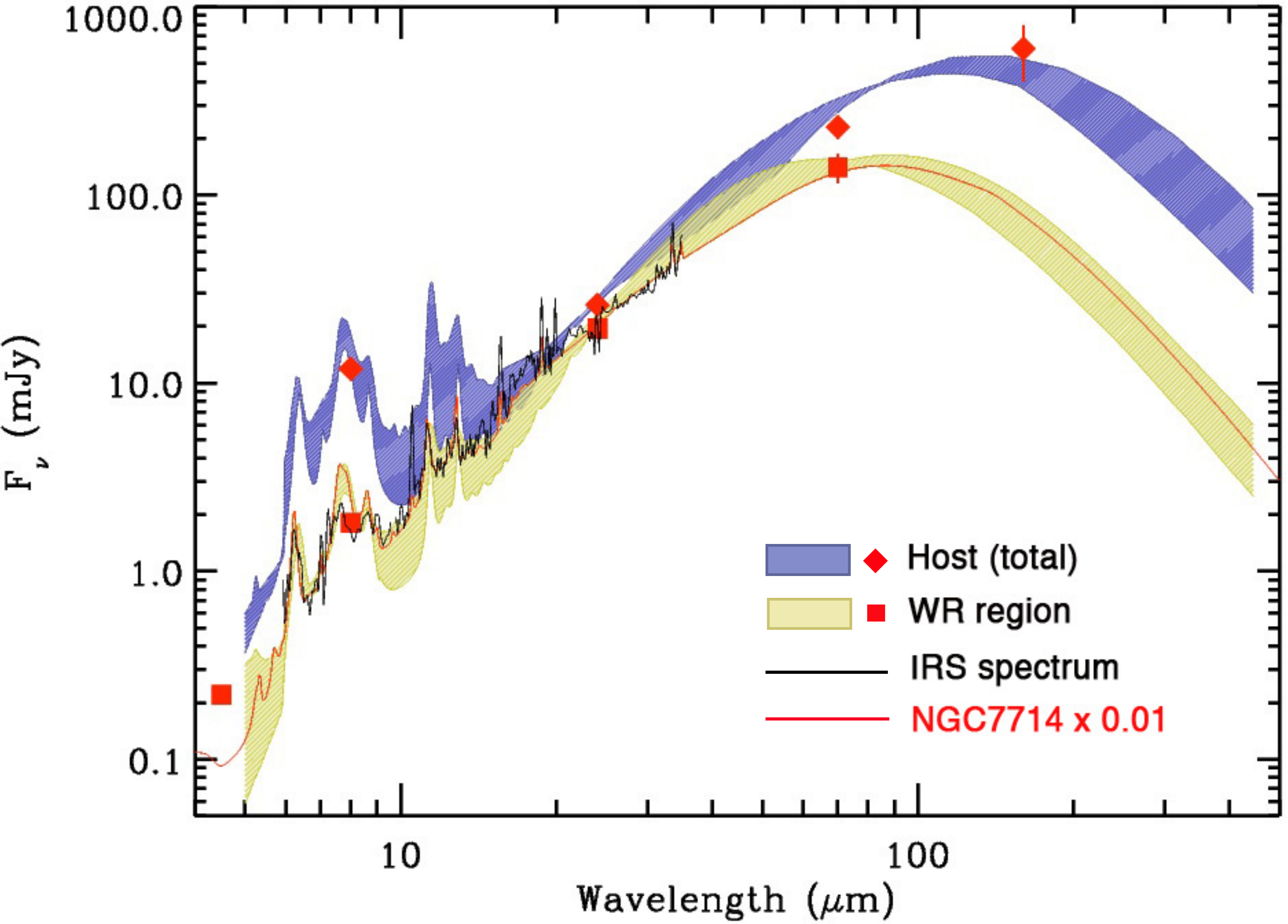}
  \caption{The \spi  broad-band photometry of the GRB\,980425 host galaxy (red filled diamonds) along with the IRAC/MIPS  photometry and the IRS spectrum of the WR~region (red filled squares, thick black line). Photometric uncertainties are shown with vertical error bars when they are larger than the symbol size. The shaded regions represent the range of possible SEDs fitting the 8--160\mic broad-band photometry of the host galaxy and the WR~region (blue and green shaded areas, respectively), defined from the 4 best-fit templates independently obtained using the libraries of \citet{Chary01}, \citet{Dale02}, \citet{Lagache04}Ê   and \citet{Siebenmorgen07} (see Sect.\,4.1 for more details).  For comparison the IR SED of NGC\,7714 taken from \citet{Marshall07} and  scaled down by a factor of 100 is also illustrated (thin red line).}
\label{fig:ir_total_sed}
\vskip .2cm
\end{figure*}

The results are illustrated in Figure\,\ref{fig:ir_total_sed}, which shows the measurements from our broad-band photometry and IRS spectroscopy together with the global range of possible SED fits
defined from the best-fit templates that were obtained for each of the 4 libraries. As expected from the different broad-band fluxes measured with IRAC and MIPS we note that the mid- to far-IR spectral slope of the WR~region is much steeper than observed for the whole GRB host.
By integrating the best fit  SEDs between 8\mic and 1000\micpa, we derived total IR luminosities of log(L$_{\rm IR}$/L$_{\odot}$)\,=\,8.66$\pm$0.04 and  log(L$_{\rm IR}$/L$_{\odot}$)\,=\,9.01$\pm$0.07 for the WR~region and the whole GRB host galaxy, respectively. In these estimates the uncertainties were obtained by combining $(i)$ the  dispersion measured between the 4 best fits obtained from the different libraries and $(ii)$ each of the dispersions measured within the individual libraries themselves. Our results thus suggest that 45$\pm$10\% of the total IR luminosity of the GRB\,980425 host galaxy originates from the WR~region close to which the GRB was observed.

The lack of photometry at submillimeter wavelengths prevents us from fully constraining the Rayleigh-Jeans emission tail, which traces the properties of the cold dust component in the GRB host galaxy and the WR~region. Therefore the position of the IR SED peak and the  characteristic temperature of the dust grain emission can not unambiguously be determined with the present data. However, since we used different libraries of templates that significantly vary from one another, the results of our SED fitting suggest that we can reasonably predict the entire IR spectral energy distribution of both the GRB host and the WR~region within an uncertainty of a factor of $\sim$\,2--3.  This can be explained by the fact that the IRAC/MIPS constraints imposed 
on the Wien side of the SED
substantially reduced the number of possible fits from those libraries.
In particular, we note that the SED of the WR~region peaks at much shorter wavelengths than the SED of the whole galaxy (as also predicted by \citealt{Michalowski09} from stellar population synthesis modeling), showing that the characteristic temperature in the WR~region is about twice larger than the average dust temperature in the interstellar medium of the galaxy. This is also a direct consequence of the larger fraction contributed by the WR~region at 24\mic than at 70\micpa.

\vskip .5cm

\subsection{SFR determination}
\label{sec:ir_sfr}

Based on  the total IR luminosities computed in the previous section we obtained a revised estimate of the star formation rates in the WR~region and the GRB host using the calibration provided by \citet{Bell05}: SFR\,(M$_{\odot}$\,yr$^{-1}$)\,=\,1.8$\times$10$^{-10}$\,(L$_{\rm IR}\,+\,$3.3\,L$_{2800}$)\,$/$\,L$_{\odot}$. In this relation, L$_{2800}$ represents the monochromatic luminosity $\nu L_{\nu}$ computed at 2800\AA, which we derived from a linear interpolation between the NUV and U-band fluxes published by \citet{Michalowski09}. In the case of the WR~region we obtained SFR\,$\sim$\,0.12\,\Msol\,yr$^{-1}$. Assuming the stellar mass M$_*$\,$\sim$\,5.7\,\,$\times$\,10$^6$\,\Msol ~ inferred by \citet{Michalowski09}, our estimate thus leads to a specific star formation rate SSFR\,$\sim$\,21\,Gyr$^{-1}$, which corresponds to a typical time scale of  $\sim$\,47\,Myr for this star-forming complex to double its  mass. 

Our new estimate of the SFR in the WR~region is smaller than the one initially published by \citet{LeFloch06} from the first \spi observations of the GRB\,980425 host (SFR\,$\sim$\,0.35\,\Msol\,yr$^{-1}$), but this first determination had relied on a large extrapolation of the 24\mic flux  to infer the total IR luminosity characterizing this environment. Our revised SFR is in good agreement with that obtained by \citet{Michalowski09} from  broad-band SED fitting (SFR\,=\,0.10\,\Msol\,yr$^{-1}$), and it is also quite consistent with the values that can be inferred   from the direct conversions recently established between the SFR and the mid-IR emission of galaxies  \citep{Alonso06,Calzetti07,Diaz_Santos08}. For example, \citet{Alonso06} obtained a tight relationship between the 24\mic and the extinction-corrected Pa\,$\alpha$ luminosities
of HII~regions and star-forming galaxies in the local Universe (SFR\,$/$\,M$_{\odot}$\,yr$^{-1}$~=~8.45\,$\times$\,10$^{-38}$~[L$_{\rm 24\mu m} /$\,erg\,s$^{-1}]$$^{0.871}$), while  \citet{Calzetti07} established a comparable correlation between  
  the SFR derived from the H$\alpha$ recombination line and the 24\mic emission of star-forming regions in nearby extragalactic sources\footnote{Note that  the relation inferred by \citet{Calzetti07} needs to be corrected by a multiplicative factor of 1.59 to match the Salpeter IMF assumed in our analysis. See the Appendix~A2 of \citet{Calzetti07} for more details.}.
  We applied these two relations to the 24\mic flux of the WR~region measured from our PSF fitting (see Sect.\,\ref{sec:mips24}), which respectively led to  star formation rate estimates of SFR\,$\sim$\,0.14\,\Msol\,yr$^{-1}$ and SFR\,$\sim$\,0.13\,\Msol\,yr$^{-1}$. Considering the possible  variations of the two calibrations with   
  the metallicity and the age of the stellar populations ($\sim$20\%, \citealt{Calzetti07}), we find therefore a very good agreement with the  revised estimate reported above  from  the UV and IR luminosities.  

In the case of the entire host galaxy, the relations from \citet{Alonso06} and \citet{Calzetti07} give a  total SFR of 0.18\,\Msol\,yr$^{-1}$ and 0.16\,\Msol\,yr$^{-1}$ respectively, while the relation obtained by \citet{Bell05} leads to a total SFR\,$\sim$\,0.66\,\Msol\,yr$^{-1}$. At optical wavelengths though, the global SED of the galaxy is dominated by the contribution of passively-evolving stellar populations \citep{Savaglio09} and 
it is likely that a large fraction of the total  UV and IR luminosities of the host is not directly produced by on-going or recent star formation. Consequently, the estimate of the SFR based on the relation from \citet{Bell05} is probably overestimated.

\vskip 1.5cm

\section{Mid-Infrared spectral properties}
\label{sec:irs_prop}

In Sect.\,\ref{sec:pahfit} we used the PAHFIT tool to  analyze the different components contributing to the IRS spectrum of the WR~region  (see Figure\,\ref{fig:pahfit}).  Focussing first on the shape of the underlying VSG~continuum, the decomposition reveals that   there is almost no absorption from the silicate features at 9.7\mic ($\tau_{9.7\mu m} \sim 0.015$). Assuming $A_V / A_{9.7\mu m} \sim 20$ as observed in the Milky Way \citep{Mathis90}, the optical depth toward the silicates correspond to a  small extinction $A_V \sim$ 0.3\,mag. This is actually consistent with the modest obscuration that was derived  toward the very center of the WR~region based on the Hydrogen Balmer decrement measured with IFU spectroscopy \citep[$E(B-V)\sim 0.03$\,mag, ][]{Christensen08}. It suggests that most of the activity in this area originates from a compact and optically-thin environment, and 
that the contribution from
dust-enshrouded star formation must be very modest. Note that our estimate appears smaller  than the value obtained by \citet{Hammer06} using optical long-slit spectroscopy ($A_V \sim 1.5$\,mag).  Such differences could originate from aperture effects, since the extinction inferred by \citet{Christensen08} reaches $E(B-V)\sim 0.17$\,mag when it is measured over the entire WR~region. It could also be due to the general difficulty in properly defining the underlying mid-IR spectral continuum around the 9.7\mic silicate feature when prominent 7.7\mic and 11.3\mic PAHs are detected. This degeneracy in the mid-IR SED fitting results in additional uncertainties affecting the determination of the obscuration from the silicates, and it could have led to a small underestimate of the extinction in the case of the WR~region.

Furthermore, the detection of prominent forbidden lines with high ionizing potential  such as [NeIII], [SIV] or [ArIII] indicates the presence of 
a particularly hard radiation field. We quantified the ionizing conditions in the WR~region  using the lines fluxes measured from both the low and high-resolution IRS spectra, which led to [NeIII]/[NeII], [ArIII]/[ArII] and [SIV]/[SIII] ratios  of 1.8, 1.05 and 1.3 respectively. Taking into account the sub-solar metallicity (Z\,$\ltapp$\,0.5\,Z$_{\odot}$) inferred by \citet{Hammer06} and \citet{Christensen08}, these ratios are comparable to what has been observed in other individually-resolved HII regions and Blue Compact Dwarves with similar chemical abundances \citep{Wu06,Gordon08,Hao09}. However these values are substantially higher than what has been typically   measured in star-forming and starburst galaxies \citep[e.g.,][]{Brandl06,Beirao08}, where the emission is likely averaged over different stellar populations spanning a wide range of properties and  ages. Assuming that the mid-IR emission of the WR~region  originates from the  central star cluster that produces most of the optical light  \citep{Hammer06}, the spatial extent of the $B$-band emission  (i.e., $\lesssim$\,100\,pc) and the hardness of the radiation field derived with IRS suggest that most of the mid-IR luminosity is produced by  very young  massive star formation. Following \citet{Rigby04b} who studied the evolution of various mid-IR line ratios with time using the CLOUDY photoionization model, we find that the [NeIII]/[NeII], [ArIII]/[ArII] and [SIV]/[SIII] ionizing indices measured in the GRB\,980425 host are  compatible with a    starburst episode younger than $\sim$\,5\,Myr. This is consistent with the estimate obtained by \citet{Hammer06} based on the number of Wolf-Rayet stars observed in this region (i.e., 1--6\,Myr).

The mid-IR emission line diagnostics that can be used to characterize the ionizing conditions of star-forming environments are obviously not independent from each other. 
A number of correlations have been established between 
[NeIII]/[NeII], [ArIII]/[ArII] and [SIV]/[SIII], as well as with [SIV]/[NeII] and with optical line ratios like [OIII]$\lambda$5007/[OII]$\lambda$3727 \citep[e.g.,][]{Giveon02,Wu06,Dale06,Groves08,Gordon08,Hao09,Bernard09}. The mid-IR spectral properties of the  WR~region are fully consistent with these different relationships, as well as with the rough correlation that exists between the metallicity of galaxies and the hardness of their radiation field  \citep{Wu06,Gordon08}. Moreover, the electron density in the WR~region can be constrained from the two Sulfur~III emission lines detected with IRS. The flux ratio [SIII]$\lambda$18.71/[SIII]$\lambda$33.53\,$\sim$\,0.8  corresponds to $N_e\sim200$\,cm$^{-3}$ \citep{Giveon02}.  This is slightly larger than the estimate that was derived from the optical spectrum of the WR~region using the [SIII]$\lambda$6716\AA /[SIII]$\lambda$6731\AA  ~ flux ratio ($N_e = 158$\,cm$^{-3}$, \citealt{Hammer06})
and it is also  higher than the densities measured in the HII~regions of M\,101 (SIII]$\lambda$18.71/[SIII]$\lambda$33.53\,$\sim$\,0.2, \citealt{Gordon08}). This could be related to the effect of the stellar winds produced by the Wolf-Rayet populations, which may have efficiently compressed the gas and the ISM of the WR~region in the host of GRB\,980425.

Finally, the equivalent widths of the PAH features characterizing the WR~region fall within expectations given the metallicity and the hardness of the radiation field previously measured.
In star-forming galaxies  the relative  strength of PAHs over the VSG continuum is known to remain constant over a wide range of environments, but it rapidly decreases with harder ionizing radiations. This is  particularly apparent when the hardness of the radiation field reaches a  threshold of [NeIII]/[NeII]\,$\sim$\,1--3 \citep{Gordon08,Lebouteiller11}, or similarly 
when the chemical abundance gets  lower than 12\,+\,log[O/H]\,$\sim$\,8--8.3 \citep[e.g.,][]{Houck04b,Engelbracht05,Wu06,Engelbracht08}. 
The properties of the  WR~region in the GRB\,980425 host galaxy  correspond to this regime of transition between extreme metal-poor regions and environments with solar metallicity, and the PAHs are clearly detected albeit with smaller EWs than observed in sources with softer radiation fields. 
Looking at Figure\,\ref{fig:plot_spec}, we also  note that the 7.7\micpa/11.3\mic  PAH inter-band ratio is  substantially smaller than observed in the mid-IR spectrum of the prototypical starburst  galaxy NGC\,7714 but it is once again in agreement with the trend that has been seen between this quantity and the ionizing index measured with  the [NeIII]/[NeII] line ratio  \citep{Smith07}. Given that the 11.3\mic feature can be produced by larger PAHs than the ones responsible for the 7.7\mic emission \citep{Draine07}, the WR~region could thus be experiencing a selective destruction of 
the small PAHs that emit at short mid-IR wavelengths \citep{Smith07}.

\vskip 0.7cm

\section{Discussion}

\subsection{Comparison with other known extragalactic sources}

 The total IR luminosity derived in Sect.\,4   (log~[L$_{\rm IR}$/L$_{\odot}$]\,$\sim$\,8.7) places the WR~region at the very bright end of the luminosity function of HII~regions observed in the local Universe \citep{Bradley06,Lee11}. Considering the physical size of the optical light distribution measured in the HST image of the host \citep[$\lesssim$\,100\,pc, ][]{Fynbo00,Hammer06},
this environment 
 is also one of  the single isolated HII~regions with the highest star formation density identified so far.
For instance, the brightest star-forming complex in the Milky Way, W49A,  releases up to L$_{\rm IR} \sim 2.7\times 10^7$\,L$_{\odot}$ within a size less than $\sim$50\,pc in diameter \citep{Ward_Thompson90}, while the total luminosity of the 30\,Doradus nebula in the Large Magelanic Cloud only reaches L$_{\rm IR} \sim 4\times 10^7$\,L$_{\odot}$ \citep{Werner78}.
Yet, 
some of the intrinsic characteristics  of this environment have already been observed in  other extragalactic sources, which suggests that such complex of star formation does not 
 represent a unique  object in the local Universe.
For instance,  isolated HII~regions with similar mid-IR spectral slope and ionization indices have been found in the giant spiral galaxy M\,101 \citep{Gordon08}.   Among these regions, NGC\,5461   does contain a population of Wolf-Rayet stars \citep{Schaerer99} and  it has a 24\mic flux of $\sim$\,920\,mJy, which given  the distance of M\,101 (6.7\,Mpc)  corresponds to a mid-IR luminosity 50\% larger than what we measured in the host of GRB\,980425. 
The mid-IR properties of the WR~region also resemble what is seen in  Blue Compact Dwarf (BCDs) galaxies with comparable oxygen abundance, 
which often exhibit a steeply rising hot dust continuum with prominent ionic lines and very little absorption by the silicate features \citep{Wu06}.
Similarly, some of the giant Super Star Clusters in the overlap region of the Antennae exhibit comparable or steeper mid-IR continuum emission as well as  higher star formation rates, although these clusters may be spatially more extended  and their radiation field is typically a bit softer  than observed in the WR~region \citep{Brandl09}.

\begin{figure}[htpb]
  \epsscale{1.2}
\plotone{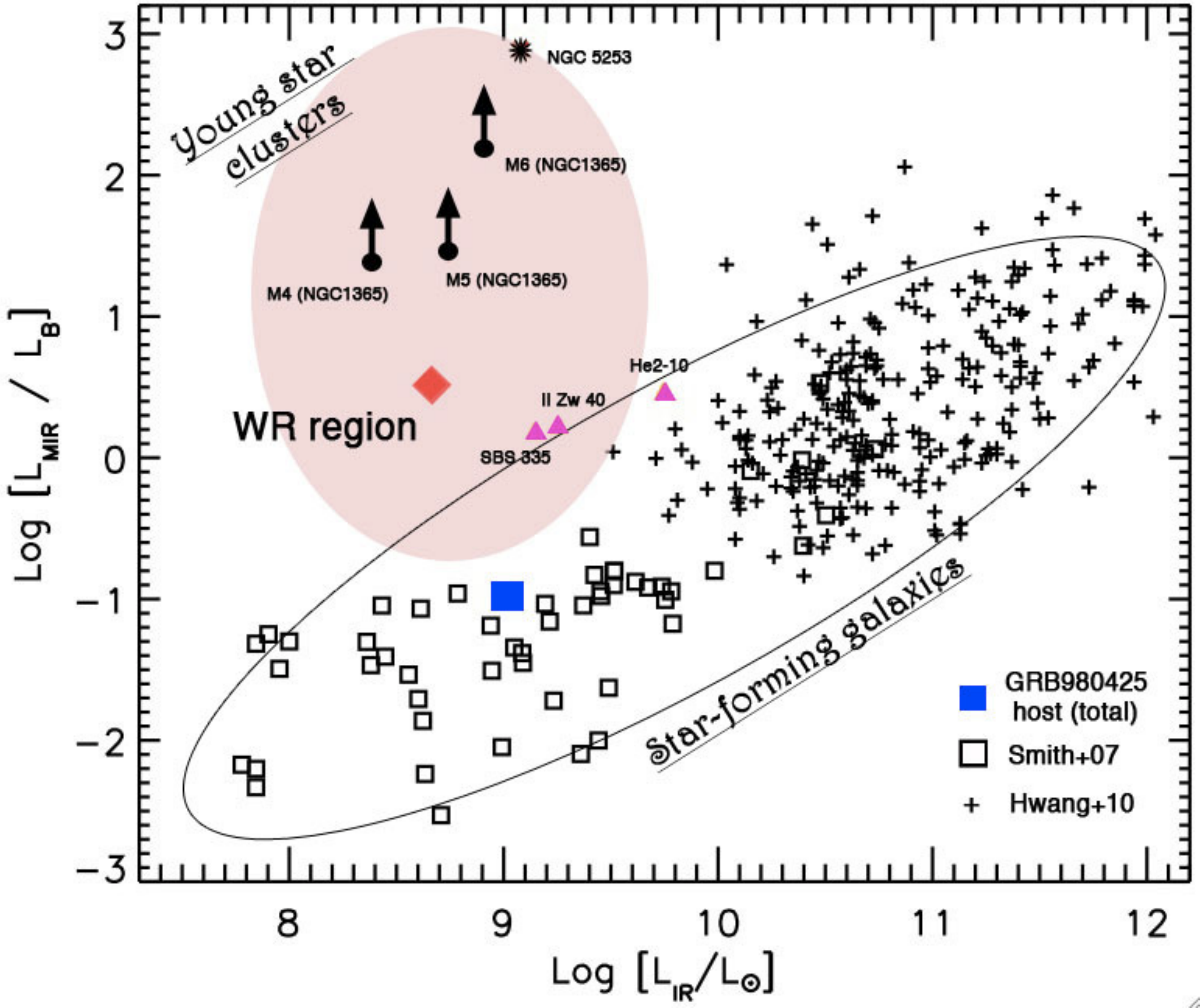}
  \caption{ The mid-IR over B-band monochromatic luminosity ratio ($L_{MIR}/L_B$)  plotted against the total IR luminosity of the WR~region (red filled diamond) and the entire GRB\,980425 host galaxy (blue filled square). For comparison we also indicate the local star-forming galaxy samples of SINGS \citep[open squares, ][]{Smith07} and SDSS  \citep['+' symbols, ][]{Hwang10}, the prototypical Blue Compact Dwarves SBS\,0335-052, II\,Zw~40 and He2-10 \citep[pink triangles, ][]{Hunt05}, the embedded starburst in NGC\,5253 \citep['$\ast$' symbol, ][]{Alonso04b} and the young star clusters of NGC\,1365 \citep[lower limits, ][]{Galliano08}. Mid-IR luminosities were computed based on MIPS-24\mic fluxes except for the SDSS sample where the IRAS-25\mic bandpass was used as a proxy.  The $L_{MIR}/L_B$ ratio in the WR~region is much higher than typically found in star-forming galaxies with comparable luminosities, but it is lower than observed in young star clusters still embedded in clouds of dust and molecular gas.
  }
\label{fig:lir_lb_comp}
\end{figure}

What is nonetheless very unusual in the host of GRB\,980425 is the large fraction (45$\pm$10\,\%) of the total Infrared luminosity of the galaxy contributed by the WR~region alone. This large contribution to the global star-forming activity of the host is not apparent at optical wavelengths, because the ratio of the mid-IR and  optical luminosities in the WR~region is much higher than measured over the whole spatial extent of the galaxy. This is illustrated in Figure\,\ref{fig:lir_lb_comp}, which shows the 24\mic over $B$-band monochromatic luminosity ratio $L_{MIR}/L_B = (\nu L_{\nu})_{24\mu m} / (\nu L_{\nu})_{B} $ of the host and the WR~region, plotted as a function of their total Infrared luminosity.  In both cases the $B$-band luminosities  were taken from  \citet{Michalowski09}. For comparison we over plotted the ratios derived from the Sloan Digital Sky Survey by \citet{Hwang10}, those measured by  \citet{Dale06} and \citet{Smith07}   in the Spitzer Infrared Nearby Galaxies Survey (SINGS), and the ratios characterizing the three BCD prototypes  SBS\,0335-052, II\,Zw\,40 and He2-10 \citep{Hunt05}. While the $L_{MIR}/L_B$  integrated over the host of GRB\,980425 is typical of quiescent star-forming galaxies with similar luminosities, we see that the ratio in the WR~region is more than an order of magnitude higher, and  comparable to that observed in Blue Compact Dwarves.

This  characteristic of the WR~region is actually consistent with what can be observed in recent starbursts, despite the apparently 
  modest extinction derived from the 9.7\mic silicate features and the Hydrogen Balmer decrement (see Sect.\,\ref{sec:irs_prop}).
 In fact, very  young and dust-embedded  stellar clusters with much higher extinction (e.g., $A_V \gtrsim 5$\,mag)
can harbor $L_{MIR}/L_B$ ratios even larger than found in the WR~region. As an example we included in Figure\,\ref{fig:lir_lb_comp} the case of the ``hypernebula" starburst enshrouded in the core of NGC\,5253 \citep{Alonso04b} as well as the massive star clusters identified in NGC\,1365 by \citet{Galliano08}. These complexes of star formation share strong similarities with the WR~region in the GRB\,980425 host in terms of SFR, age and luminosities. Yet, the optical depth toward these sources is much more significant, explaining their extreme optical-to-infrared colors. Assuming that the obscuration and the IR/optical luminosity ratio decrease with the starburst age, what we are seeing in the WR~region could thus be one or several star clusters that have already evolved and partly escaped the original molecular cloud where they were born.  Alternatively,   and rather than an ``aging" effect, the difference of visual extinction between the WR~region of the GRB\,980425 host  and the enshrouded clusters of NGC\,5253 and NGC\,1365 
 could also be a direct consequence of the presence of the WR~stars themselves, which may have accelerated the process of   clearing out the local ISM of the star-forming region via their emission of long-lasting supersonic winds. 
Depending on geometry effects such a scenario could lead to small silicate absorption as observed, while a still important emission of very hot dust at mid-IR wavelengths could arise from amorphous carbon grains heated to very high temperatures by the massive stars dominating the spectrum of the WR~region.

\vskip 1cm

\subsection{Long Gamma-Ray Bursts and super-luminous HII~regions}

 As demonstrated in Sect.\,4, the most striking result of our analysis resides in the bolometric luminosity of the GRB\,980425 host, half of which  originates from a single and compact star-forming complex located at more than 2\,kpc from the galaxy center. 
This large contribution of extra-nuclear star formation has also been observed in a  few other spiral galaxies of the local Universe \citep[e.g., NGC\,5257, NGC\,6670, Arp\,256,][]{Haan11}, although these cases are rare and they have much higher mass and SFR than the host of GRB\,980425. Most often,
  the star-forming activity of morphologically-evolved systems in the luminosity range of LGRB hosts appears to be preferentially distributed among a large number of HII~regions or centrally concentrated toward ring-like structures of multiple knots and star clusters.
The fact that this atypical configuration is found in a galaxy initially selected as the host of a rare event such as a GRB raises therefore the question of a possible link between the physical properties of  the WR~region and the trigger of the long Gamma-Ray Bursts in star-forming galaxies.

Indeed, the  properties observed in this environment look similar to a number of specific characteristics commonly found in GRB host galaxies at cosmological distances. First, the mid-IR spectroscopic features and the SSFR of the WR~region both confirm the presence of extremely young star formation (i.e., $<$\,5\,Myr) in this area. Assuming a constant star formation rate, the SSFR that we derived in Sect.\,\ref{sec:ir_sfr} indicates that this complex of massive star formation would double its stellar mass in only 47\,Myr. This is extremely short  compared to the typical time scale of galaxy  evolution, 
and although it is not as extreme in other GRB host galaxies it is consistent with the higher SSFR displayed by this population when compared to field star-forming galaxies at similar redshifts
\citep{Christensen04,Thoene08b}. Second, the WR~region has sub-solar metallicity, only modest extinction by dust and therefore very blue optical colors \citep{Fynbo00,Hammer06,Christensen08,Michalowski09} similar to what is typically observed for GRB hosts at higher redshifts \citep{Fruchter99,LeFloch03,Han10}.
Among the numerous HII~environments detected in the GRB\,980425 host it also harbors the largest surface brightness region of the galaxy, which is consistent with the link that was established by \citet{Fruchter06} between the location of GRBs and the very brightest regions of the hosts. Finally, optical emission line ratios like [OIII]/H$\beta$, [NII]/H$\alpha$, [SII]/H$\alpha$ and [NII]/[OII] suggest that the ionizing conditions and the stellar populations in the WR~region are very similar to that observed in more distant and unresolved GRB host galaxies \citep{Christensen08}.  Without claiming a direct link between the birth of the GRB\,980425 progenitor and the WR~region, it would seem therefore difficult from all of these observations to argue {\it against\,} a kind of causal link between the two.

 Besides, and contrary to the characteristics of  the WR~region, it is worth noting  that the integrated properties of the GRB\,980425 host galaxy differ from those of other GRB hosts at high redshift. For example, the host of  GRB\,980425 is an already-evolved galaxy that is still forming stars, but at a very low rate with respect to the stellar mass that it has already built \citep{Savaglio09,Michalowski09}.  Furthermore the [OIII]/H$\beta$ line ratio measured in the other HII~regions of the GRB\,980425 host are  substantially smaller than what is typically observed in more distant LGRB host galaxies \citep{Christensen08}. These differences further emphasize the connection that must exist between the episode of star formation currently taking place in the WR~region and the trigger of the GRB\,980425 in this galaxy. 

Nonetheless, the most unsettling fact is that  GRB\,980425 {\it did not\,} occur within the WR~region but at a projected distance of $\sim$\,900\,pc away.  What can then be the physical connection between the two\,? Long GRBs represent extremely rare phenomena which are more likely to occur from large populations of massive stars. Given the very low  density of the latter  in the area where GRB\,980425 was found, and considering
the presence of numerous Wolf-Rayet and OB stars in the WR~region, 
\citet{Hammer06}  suggested a ``run-away scenario'' in which the GRB\,980425 progenitor may have come from the WR~region itself. This progenitor would have been dynamically ejected after a kick or close stellar encounters, and traveled 
 at velocities $\sim$\,200--300\,km\,s$^{-1}$ across the ISM of the galaxy before ending its life as a GRB. Very massive runaway stars with such velocities are actually expected from numerical simulations of dynamical encounters in stellar clusters \citep[e.g.,][]{Gvaramadze11}. In the mid-infrared, the detection of very high-excitation transition lines such as  [OIV] at 25.9\mic can indicate the presence of  very hot Wolf-Rayet stars \citep{Schaerer99b} or recent supernovae, which in the case of  the WR~region could further  strengthen this possible connection with the progenitor of GRB\,980425.
In the IRS spectrum however, the emission from [OIV] is only observed at the 2$\sigma$ level,  making this association uncertain.
 
   Conversely, star formation is still on-going at the location where GRB\,98425 was observed and therefore one can not fully rule out that its progenitor was truly born in this area. For instance,  this site could have experienced several million years ago a massive starbursting episode strongly biased toward the high mass end of the IMF, hence explaining the relatively low density of 
stars now observed in this area compared to that of the WR~region. In this second scenario, it is the physical mechanism that triggered star formation at both locations, as well as similarities between the chemical properties of these two environments,
that would explain a possible  relationship between the WR~region and  GRB\,980425.  For instance, a single density wave propagating through the disc of the galaxy could have triggered over a short time scale but in a sequential way the formation of new stellar populations along the south-eastern spiral arm, and we would  be seeing the host while the star-forming episode at the  GRB location has just terminated.
In fact internal processes must have played a critical role in driving the recent star-forming activity in this object. The galaxy  ESO\,184-G82 appears as an isolated system in the sky \citep{Foley06} and a trigger of star formation due to the accretion of external material via minor merging seems quite unlikely.

\subsection{Perspectives}
 
The current data are unfortunately  not sufficient  to  determine the exact origin of the progenitor of GRB\,980425. Further progress may be expected from IFU kinematic studies, which could provide better
constraints on the velocity fields and the dynamical properties of the stellar populations in the host galaxy. High-resolution maps of the molecular gas distribution inferred from facilities like ALMA would  also help constraining how the star formation efficiency varies in the  surrounding of the WR~region.

Similarly, characterizing with more statistically representative samples the locations of GRBs with respect to the distribution of star-forming regions in their hosts 
should provide more insights into the connection between the occurrence of LGRBs and the dynamical ejection of massive stars in stellar clusters. A clear trend has  already been observed by \citet{Fruchter06} between the locations of cosmological GRBs and the brightest regions of their hosts \citep[see also ][]{Svensson10}. However the spatial resolution of current facilities has not yet been sufficient to constrain possible offsets at sub-kpc scales. Further improvements should thus be expected from
 next generation instruments like $JWST$ and the {\it Extremely Large Telescopes}. 

Independently, the properties of the WR~region in the host of GRB\,980425 reinforce the idea that LGRBs are  triggered {\it within or in the vicinity\,} of extremely young but active regions of massive star formation, lending further support to their collapsar origin \citep{MacFadyen99}. Similar to what can be inferred from the chemical abundance of the WR~region \citep{Hammer06,Christensen08} this star-forming activity tends to be associated with metal-poor environments, hence explaining that LGRBs rather occur within relatively unobscured blue galaxies with sub-solar oxygen abundances \citep{Han10}.

\vskip 1cm

\section{Summary and Conclusions}

Thanks to its relatively small distance (i.e., 36\,Mpc, \citealt{Galama98}) the host of GRB\,980425 represents one of the very few cases of LGRB hosts where spectroscopic and bolometric properties can be spatially resolved. It thus offers a unique opportunity  to constrain  the characteristics of the close environment where LGRBs take place. In this work we present \spi 5--160\mic spectro-imaging data of the GRB\,980425 host galaxy and the ``Wolf-Rayet region" close to which the GRB was observed. Our main results can be summarized as follows:

\begin{itemize}
\item{}Given the compactness of the WR~region ($<$\,100\,pc in diameter), the highly-ionized emission lines in our IRS spectrum (e.g., [NeIII],  [SIV], [ArIII])
 suggest that this environment
 is  associated with  extremely young ($<$\,5\,Myrs) 
 massive star formation. 

\item{}While the whole GRB\,980425 host galaxy is producing stars at a very low rate compared to its already-formed stellar populations, the WR~region will double its stellar mass in only 47\,Myrs assuming constant SFR. It has a total IR luminosity of $\sim$\,4.6\,$\times$\,10$^8$\,L$_{\odot}$ and it represents one of the most luminous HII~regions  identified in the nearby Universe.

\item{} In the WR~region the relative strength of the PAH features over the continuum is similar to what has been observed in other systems with comparable oxygen abundance (Z\,$\ltapp$\,0.5\,Z$_{\odot}$).  The extinction measured from the optical depth of the 9.7\mic silicate feature implies  modest obscuration ($\tau_{9.7\mu m} \sim 0.015$).

\item{} Although the WR~region produces  less than 5\% of the $B$-band emission of the GRB host, it contributes up to 
45$\pm$10\,\% of the total IR luminosity of the galaxy. Its IR SED also peaks at much shorter wavelengths than the global SED of the host, suggesting that the characteristic temperature in the WR~region is about twice that of the average dust temperature in the interstellar medium of the system.
 
\item{}  The ratio between the mid-IR and the $B$-band luminosities in the WR~region is characteristic of recent starbursts and much higher than what is observed in star-forming galaxies with similar L$_{\rm IR}$.
 However it is not as extreme as  in very young and dust-embedded  stellar clusters that harbor much stronger extinction than derived from the mid-IR spectrum of the GRB~host. Given the blue  colors of the WR~region at optical wavelengths, this suggests that this complex of star formation  harbors
  one or several star clusters that have already evolved and partly escaped the original molecular cloud where they were born.

\end{itemize}

 The contribution of the Wolf-Rayet region to the bolometric output of  the  GRB\,980425 host makes  this galaxy a rather unique object of the nearby Universe. Considering that GRBs represent extremely rare events, our analysis  supports therefore the existence of a causal link between the origin of GRB\,980425 and the activity of star formation in the WR~region. More generally, and taking also account of the properties characterizing the whole population of  LGRB hosts, our results corroborate the common idea that the triggering of long GRBs is mostly  associated with   very active star formation in chemically-young  environments.

\acknowledgments This work was enabled based on the funding from the
IRS and the MIPS projects which are supported by NASA through the Jet
Propulsion Laboratory (subcontracts \#1257184 \& \#960785), and thanks to
 the efficient technical support provided by the {\it Spitzer\,} Science Center.  We want to thank our referee for relevant suggestions and a careful review of the manuscript, as well as Michal Micha{\l}owski for useful comments on our work. We are particularly grateful to Vianney Lebouteiller for fruitful discussions
 and for providing us with updated IRS data reduction prior to the publication of the Cornell AtlaS of Spitzer IRS Sources \citep[CASSIS,][]{Lebouteiller11b}, and to Ho~Seong Hwang for sharing his results on the SDSS data.
 We also want to acknowledge Yanling Wu  for  her help in the IRS data reduction,  as well as Jason Marshall for providing  some of his IR SED modeling and  JD~Smith for making publicly  available his PAHFIT routines. 
 We greatly appreciated the help from Stephane Arnouts and Olivier Ilbert when using the code $LePhare$.
 We finally thank Tanio D{\'{\i}}az-Santos, Fr\'ed\'eric Galliano, David Elbaz, Suzanne Madden and Marc Sauvage for useful discussions.

\end{document}